\documentclass[aps,twocolumn]{revtex4}
\usepackage{graphics}
\usepackage{epsfig}
\usepackage{amsmath}
\usepackage{color}
\usepackage{amssymb}

\newcommand{\Ket}[1]{|#1\rangle}
\newcommand{\Bra}[1]{\langle#1|}

\newcommand{\la}{-} %{\leftarrow}
\newcommand{\ra}{+} %{\rightarrow}
\newcommand{\eins}{\ensuremath{\mathbb{I}}}
\newcommand{\hil}{\ensuremath{\mathcal H}}

\newcommand{\NOT}{X_\textrm{\textsc{\uppercase{NOT}}}}
\newcommand{\NOTs}{X'_\textrm{\textsc{\uppercase{NOT}}}}
\newcommand{\PHASE}{X_\textrm{\textsc{\uppercase{PHASE}}}}
\newcommand{\Hop}{\ensuremath{\textrm{\textsc{\uppercase{H}}}}}
\newcommand{\ew}[1]{\ensuremath{\langle #1 \rangle}}
\newcommand{\Span}[1]{\text{span}\left\{#1\right\}}

\begin{document}
%\title{{\bf Quantum walks in one and two dimensions in arrays of optical traps}}
\title{{\bf One- and two-dimensional quantum walks in arrays of optical traps}}
\author{K.~Eckert$^{1}$, J.~Mompart$^{2}$, 
G. Birkl$^{3}$, and M. Lewenstein$^{1,4,}$\footnote{also at Instituci\'o
Catalana de Recerca i Estudis Avan\c cats.}}
\affiliation{ \mbox{$^{1}$}Institut f{\"u}r Theoretische Physik, Universit\"at Hannover, D-30167 Hannover, Germany.\\
\mbox{$^{2}$}Departament de F\'{\i}sica, Universitat Aut\`onoma de Barcelona,  E-08193 Bellaterra, Spain. \\
\mbox{$^{3}$}Institut f\"ur Angewandte Physik, Technische Universit\"at Darmstadt, Schlossgartenstra{\ss}e 7, D-64289 Darmstadt, Germany.\\
\mbox{$^{4}$}ICFO - Institut de Ci\`encies Fot\`oniques, 08034 Barcelona, Spain.}\date{\today}

\begin{abstract}

We propose a novel implementation of discrete time quantum walks for a neutral atom in an array of optical microtraps or an optical lattice.
We analyze a one-dimensional walk in position space, with the coin, the additional qubit degree of freedom that controls the displacement of the quantum walker,
implemented as a spatially delocalized qubit, i.e., the coin is also encoded in position space. We analyze the dependence of the quantum walk
on temperature and experimental imperfections as shaking in the trap positions.
Finally, combining a spatially delocalized qubit and a hyperfine qubit, we also give a scheme to realize a quantum walk on a
two-dimensional square lattice with the possibility of implementing different coin operators.


\end{abstract}

\maketitle

%Standard Deviation -> Variance: $\sigma$ -> $\sigma^2$

\section{Introduction}

In classical computation, random walks are powerful tools to address a large number of problems in many areas of science,
as, for example, graph-connectivity or satisfiability problems \cite{class_alg}. 
It is this success of random walks that motivated to study their quantum analogues in order to
explore whether they might extend the set of quantum algorithms. Two distinct types of quantum walks have
been identified: for the {\it continuous time quantum walk} a time-independent Hamiltonian 
governs a continuous evolution of single particle in a Hilbert space spanned by the vertices of a graph \cite{contQWalks},
while the {\it discrete time quantum walk} requires a {\it quantum coin} as an additional degree of freedom in order to  
allow for a discrete time unitary evolution in the space of the nodes of a graph. The connection between both types
of quantum walks is not clear up to now \cite{KempeReview}, but in both cases different topologies of the underlying graph
have been studied, e.g., discrete time quantum walks on circles \cite{QWinCircle}, on an infinite line \cite{QWonLine},
on more-dimensional regular grids \cite{QWmoreDim}, and on hypercubes \cite{QWonHC}.
The field has recently been reviewed by Kempe \cite{KempeReview}.

Several algorithms based on quantum walks have been proposed \cite{qw_algo, qw_algo2, qw_algo3,coinsfaster}.
To implement such an algorithm in a physical system
it ultimately has to be broken down into a series of gates acting on a register of qubits \cite{KempeReview}. From the more fundamental
point of view however more straight-forward implementations are interesting, i.e., direct implementations
of a quantum walker (a particle, a photon etc.) moving, e.g., in position or momentum space. So far some setups for one-dimensional 
realizations have been analyzed, including trapped ions \cite{QWIonTrap},
neutral atoms in optical lattices with state dependent potentials \cite{QWOptLat},
single-photon sources together with linear optical elements \cite{QWLOE}, and also with classical optics \cite{knight}.
%Here we propose 
%%to implement discrete quantum walks in a arrays of optical microtraps \cite{PRLBirkl}. We show 
%different setups with neutral atoms to realize the walk on the line, where the coin degree of freedom is not given by some internal
%levels of the atom, but is realized in space. We use the idea of spatially delocalized qubits developed in
%\cite{PRLJordi}, such that {\it only} a spatial variation of the potentials is necessary, without the need for state dependent
%potentials or additional external lasers. The qubit is implemented in the presence of the atom in the ground state of one
%out of two trapping potentials.
Here we use the idea of spatially delocalized qubits (SDQ) developed in \cite{PRLJordi} to propose a novel quantum walk implementation
with neutral atoms. The particle is walking in position space, but in contrast to the proposal in \cite{QWOptLat} also the quantum coin
is represented by a spatial degree of freedom, as it is implemented by the presence of the atom in the ground state of one out of
two trapping potentials. The particle is manipulated only by varying the trapping potentials, which induces tunneling between traps,
and no state dependent potentials are necessary.
This concept can be applied to neutral atoms trapped in optical lattices  \cite{opt_lat}, in magnetic potentials \cite{mag_pot}, 
as well as in arrays of microtraps \cite{PRLBirkl}; here we will
especially analyze the latter case. We will also show how a combination of a spatially delocalized qubit and a hyperfine qubit
together with state dependent potentials allows to implement a quantum walk on a two-dimensional square lattice. 
Quantum walks in higher dimensions offer a very rich structure of dynamics, and recently a spatial search algorithm 
using a modified quantum walk on a two-dimensional grid has been proposed \cite{coinsfaster}.

For the one-dimensional case we will discuss the influence of non-adiabatic processes and of shaking of the
trap positions, and we will estimate the effect of decoherence. We will also consider dependencies of the quantum walk
on the vibrational trapping state and thus on the temperature and show that, within a range of parameters accessible
in experiments, a transition from the quantum walk to the classical random walk can be studied. This is not only interesting
from a fundamental point but also allows to assess the degree of control that can be reached in the experiment.

It has been noted \cite{classical,knight,classical2d}, that essentially only interference is necessary for a quantum walk, such that it can be implemented
with classical fields. Nevertheless considering setups with neutral atoms is justified by a strong interest in these
systems as tools for quantum computation \cite{qc}, as well as by the possibility to include further effects as, e.g., quantum walks with two
or more (possibly interacting) particles \cite{yasser}.

\section{Quantum walks and optical microtraps}
\setcounter{subsubsection}{0}

\subsubsection{Quantum walks}

Let a particle move on a one-dimensional infinite line, such that it can only hop between discrete sites
$x=k\cdot a$ labeled by $k\in\{\ldots,-2,-1,0,1,2,\ldots\}$,
with $a$ being the distance between sites. At each 
time step the particle moves with equal probability to either of the adjacent sites.
For a classical random walk, the probability for the particle to 
be at a certain site for a large number of steps approaches a gaussian function centered around its initial position $x_0$, 
with the variance
$\sigma^2=\ew{(x-x_0)^2}$ growing linearly with the number of steps $n$.
For the quantum version a state $\Ket{k}$ is attached to each site $x=k\cdot a$, i.e., the particle
is walking in $\hil_W=\Span{\Ket{k},k=\ldots,-2,-1,0,1,2,\ldots}$. However, the random move cannot be just
replaced by walking to the left and to the right in superposition, as this turns out to be non-unitary \cite{meyer}.
For this reason a {\it quantum coin} is introduced as an additional degree of freedom. In the simplest case of the
quantum walk on a line, the coin space $\hil_C$ is two-dimensional and we will denote the states that span 
$\hil_C$ by $\Ket{\la}$ and $\Ket{\ra}$,
and the total Hilbert space is thus $\hil_W\otimes\hil_C$. Each step of the quantum walk is then composed from
two operations: (i) applying a unitary operation $C$ to the coin (simultaneously at all sites), e.g., a Hadamard
operation $C=\Hop$:
%(which except for a phase is the unique unbiased coin for the walk on the line \cite{coinsandstates}),
\begin{equation}
(\eins\otimes \Hop)\Ket{k,\pm}=\frac1{\sqrt{2}}(\Ket{k,+}\pm\Ket{k,-})\quad\forall k,\label{h1d}
\end{equation}
followed by (ii) applying a displacement operation $O_{\text{1D}}$ which moves the particle left or right depending on the coin:
\begin{equation}
O_{\text{1D}}\Ket{k,\pm}=\Ket{k\pm1,\pm}\quad\forall k,\label{o1d}
\end{equation}
where we have not explicitly written the tensor product: $\Ket{k}\otimes\Ket{\pm}\equiv\Ket{k,\pm}$ etc.
The probability distribution arising from the iterated application of $W=O_{\text{1D}}(\eins\otimes \Hop)$ is, except for
the first three steps, significantly different from the distribution of the classical walk: 
if the coin
initially is in a suitable superposition of $\Ket{\la}$ and $\Ket{\ra}$ it has two maxima symmetrically displaced from
the starting point. In general the exact form of the distribution, especially the relative height
of the maxima, depends on the initial coin state.
Compared to the classical random walk its quantum version propagates faster along the line:
its variance grows quadratically with the number of steps $n$, $\sigma^2\propto n^2$, compared to $\sigma^2\propto n$
for the classical random walk.

For the walk on a line, $\Hop$ is, up to phases (which can be absorbed also into the initial state), the only unbiased coin operator \cite{coinsandstates}.
For a two-dimensional regular square lattice a much richer structure of
coin operators and possible probability distributions arises.
As has been observed by Mackay {\it et al.} \cite{QWmoreDim} and by Tregenna {\it et al.} \cite{coinsandstates}
in this case different unbiased coin operators and initial states can be chosen that produce significantly different dynamics,
ranging from distributions with a sharp centered spike to distributions having the shape of a ring.
%which is
%he key feature for the speed-up of algorithms based on quantum walks \cite{KempeReview,expspeedup}.
%The quantum walk on the circle is described by the same 
%operators, eqn. (\ref{h1d}, \ref{o1d}), but here $\hil_W=\Span{\Ket{k},k=0,1,2,\ldots K}$ has finite dimension and is cyclic,
%{i.e} $\Ket{K+1}$ is identified with $\Ket{0}$. In this case the distinguishing feature from the classical walk is the mixing
%time, {i.e.} the number of steps in which the over all earlier steps averaged distribution becomes unity \cite{QWinCircle}.
%Finally, for the quantum walk on the infinite two-dimensional plane the Hilbert space $\hil_W$ is spanned by state $\Ket{k,l}$

\subsubsection{Optical microtraps}\label{expsetup}

As a particular setup for the implementation, we consider the controlled motion of neutral atoms in arrays of optical microtraps.
The microtraps are created by illuminating a set of microlenses with a red detuned laser beam, such that in each of the foci of the individual lenses neutral
atoms can be stored 
by the dipole force \cite{PRLBirkl}. By illuminating the set of microlenses by two independent laser beams, it is possible
to generate two sets of traps which can be approached or separated by changing the angle between the two lasers. This allows the
atom to propagate between different microtraps. The optical potentials have a gaussian shape, i.e.,
\begin{equation}
V(x)=-V_0\exp\left({-\frac1{2V_0}m\omega_x^2x^2}\right)=-V_0\exp\left({-\frac{\hbar\omega_x}{2V_0}(\alpha x)^2}\right).
\label{eq:pot}
\end{equation}
For the simulations we present here, we will use $V_0=200\;\hbar\omega_x$. In this case the traps are deep enough to be
described by harmonic potentials of frequency $\omega_x$ in the limit of large separation.
Then $\alpha^{-1}=\sqrt{\hbar/m\omega_x}$ denotes the spread of the ground
state in position space, with $m$ being the mass of the atom.

For the preparation of the initial state we assume that a single atom can be placed in the ground state of a specific trap. 
For this reason, and also to be able to read out the final state of the system, it is necessary to be able to address each trap separately.
This addressability has already been demonstrated in arrays of up to $10\times10$ microtraps \cite{PRLBirkl}.

\section{One-dimensional walks}

\setcounter{subsubsection}{0}

The implementation of the coin at each site $k$ will follow the idea of spatially
delocalized qubits from \cite{PRLJordi}, i.e., the basis states $\Ket{k,\pm}$ will
be represented by a single atom occupying the ground state of one of two adjacent traps.
Unitary operations are performed by approaching the two traps
forming the coin, allowing the atom to tunnel between them. In the following we will use
quantum optics notation to describe the effect of tunneling between traps, e.g., an
operation exchanging the population of two traps will be termed $\pi$ pulse and
a Hadamard-like operation $\Ket{k,\pm}\mapsto\frac1{\sqrt2}(\Ket{k,+}\pm i\Ket{k,-})$
will be termed $\pi/2$ pulse \cite{PRLJordi}.

\begin{figure}[tbph]
\begin{center}
\includegraphics[width=\columnwidth]{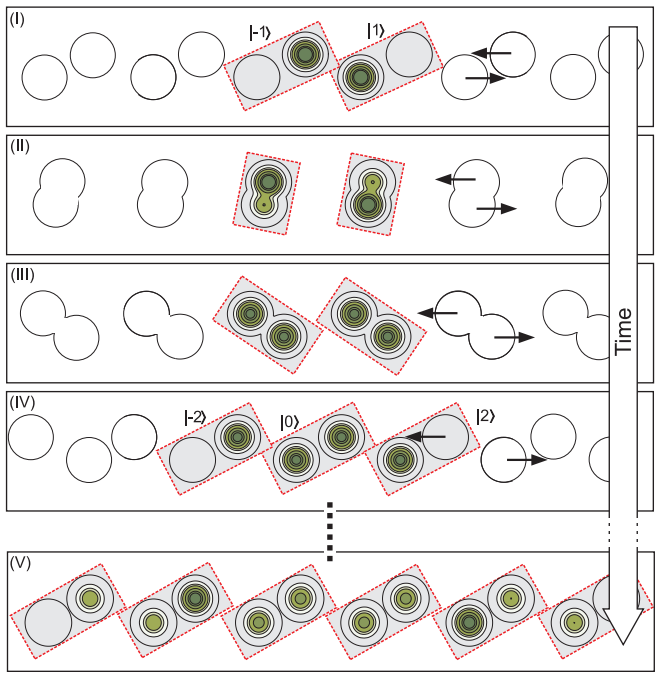}
\end{center}
\caption[]{(Color online)
Configuration with two rows of traps, the qubit is implemented 'perpendicular' to the rows (dashed rectangles show which two traps form
each qubit). The upper (lower) row moves left (right) with constant velocity (see arrows on right). (I) After the first step:
$\Ket{\psi}=1/\sqrt{2}(\Ket{-1,-}+\Ket{+1,+})$;
(II)-(III) the coin operation, in this case a Hadamard gate, is performed when the traps pass each other; (IV) the shift $O_{1D}$
is implicit through a redefinition of the qubits. After an even (odd) number of shift operations only the even (odd) qubits
(compared to the standard quantum walk definition) are defined;
(V) the probability distribution after the sixth displacement operation.
For the numerical simulation
we used a potential which, along the line connecting the centers of two traps, reads
$V(x)=\hbar\omega_x\min\left\{\alpha^2(x-a)^2,\alpha^2(x+a)^2\right\}$.
The velocity is chosen such that during
the passing of two traps a Hadamard operation is performed. The initial state is 
$\Ket{\psi_{\text{init}}}=\frac{1}{\sqrt{2}}\left(\Ket{0,\la}+\Ket{0,\ra}\right)$.
} 
\label{fig_conf1}
\end{figure}
We propose two closely related configurations, both leading
to a quantum walk.
For the first configuration two rows of traps are necessary. Each coin is defined through one trap from each row.
By moving both rows in opposite directions with appropriately chosen distance and velocity, the coin operations are performed 
when the traps pass each other at close distance. The displacement is implicit through a redefinition of the coin each time two traps
have passed. Fig.~\ref{fig_conf1} shows the first steps in the temporal evolution
of the configuration with two rows of traps along with the corresponding
probability distributions resulting from an integration of the two-dimensional
Schr\"odinger equation (see figure caption for details).
Fig.~\ref{fig_conf1} (I--III) show the coin operation, (III-IV) the redefinition of the coins and (V) shows the probability distribution
after the sixth displacement operation.
The onset of the quantum walk character of the distribution is clearly visible as two maxima symmetrically displaced from the origin appear.
%Note that due to the redefinition of the coins there are no empty sites, as it is 
%the case for the quantum walk defined above, where at even (odd) times only even (odd) sites are occupied.
If the continuous displacement of one row with respect to each other requires mechanical movement of an array of lenses
this setup is quite challenging, a problem which might be overcome by using holographic techniques to generate arrays of microtraps
\cite{bargamini}.

\begin{figure}[tbph]
\begin{center}
\includegraphics[width=\columnwidth]{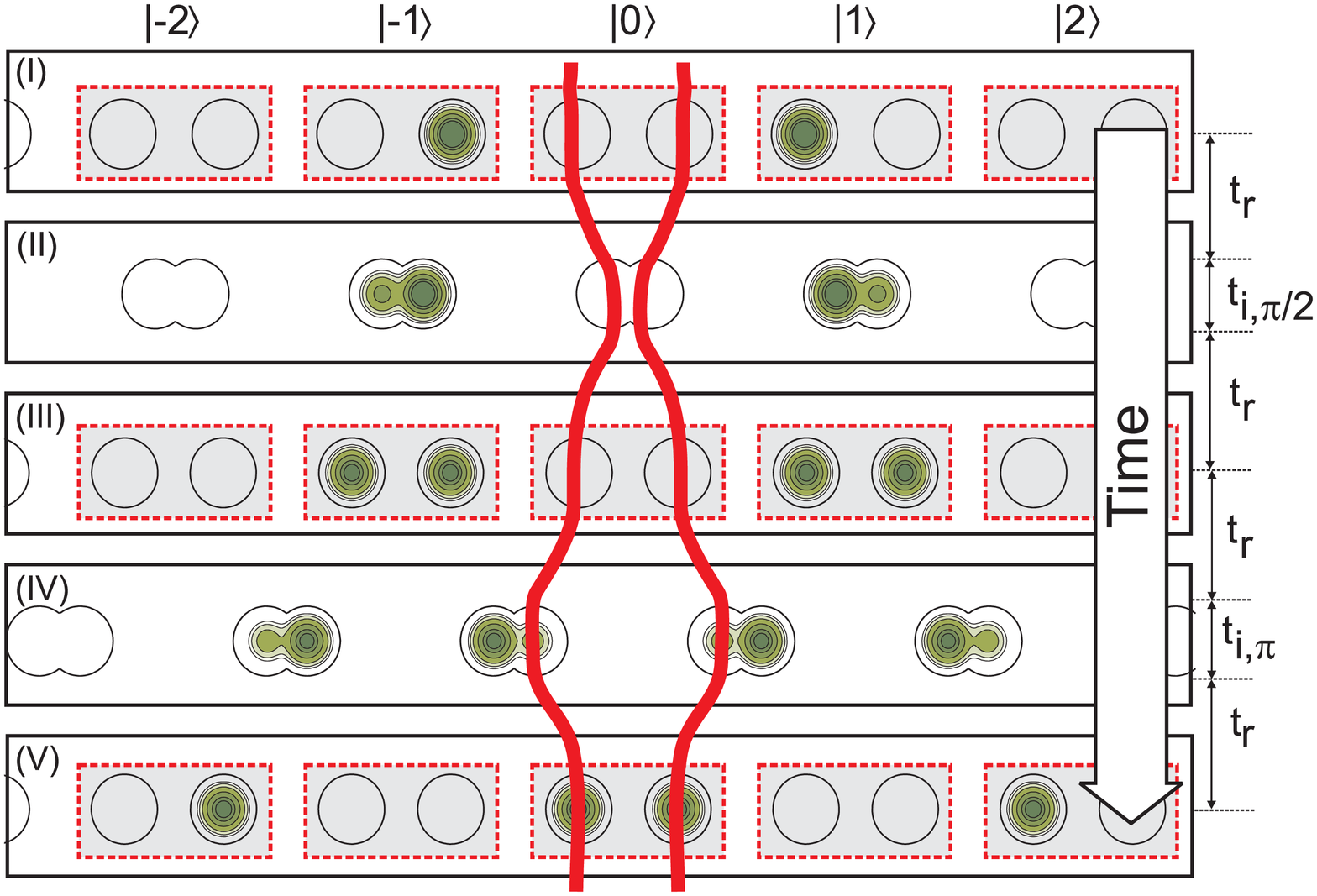}
\end{center}
\caption[]{(Color online)
Configuration with one row of traps, the qubit is implemented 'parallel' to the rows (grey boxes). (I) After the first step:
$\Ket{\psi}=(\Ket{-1,+}+\Ket{+1,-})/\sqrt{2}$;
(II) traps inside each qubit are approached to give the coin operation; (IV) the shift $O_{\text{1D}}$
is realized through approaching traps of adjacent qubits.
}
\label{fig_conf2}
\end{figure}

In what follows we will concentrate on the coin being implemented 'parallel' to the direction of displacement, such that only a single
line of traps is necessary, see Fig.~\ref{fig_conf2}.  Labeling
the traps of the $k$th qubit by $2k$ and $2k+1$, for coin operations the traps $2k$ and $2k+1$ are approached, 
while for the steps in the walk a $\pi$-pulse between traps in adjacent qubits, i.e., between
traps $2k+1$ and $2(k+1)$, moves the atom one step to the left or to the right, respectively. Contrary to the displacement operator from
Eq.~(\ref{o1d}), this procedure flips the coin operator at each move, i.e., we have 
$O_{\text{1D}}\Ket{k,\pm}=\Ket{k\pm1,\mp}$ (termed flip-flop walk in \cite{coinsfaster}).
Clearly the experimental requirement is to be able to move all odd (or all even) traps as a whole
to both directions, thus approaching each second trap to its left or right neighbor. This can be realized as described in section
\ref{expsetup} in optical microtraps \cite{PRLBirkl}, but also in optical lattices \cite{vibrational,weitz1} or magnetic microtraps \cite{mag_pot}.

For the gaussian trapping potentials of Eq. (\ref{eq:pot}), Fig.~\ref{fig_1d_gs} (i) shows
a simulation of the quantum walk for potential depth $V_0=200\;\hbar\omega_x$ and an initial separation
$\alpha a_{\text{max}}=60$, obtained from an integration of the one-dimensional Schr{\"o}dinger equation
using Fourier transformation and a split-step method. Initially the atom is prepared in an equal superposition
of the two ground states of the central qubit, i.e., of the two central traps, such that
$\Ket{\psi_{\text{init}}}=\frac{1}{\sqrt{2}}(\Ket{0,+}+\Ket{0,-})$.
The distance of the traps is changed between the maximal value $\alpha a_{\text{max}}=60$ and a minimal value
$\alpha a_{\text{min}}=28.8$. The latter distance is for the given trapping parameters close enough for tunneling to take place. 
Moving the traps adiabatically between this distances requires 
techniques to optimize the moving process while suppressing transitions between motional states \cite{optimization}. In this way,
the time $t_r$ necessary to approach -- or separate -- the traps can be reduced to $\omega_xt_r=100$ or below while maintaining
a fidelity larger than $F=0.999$ \cite{PRLJordi}.
The time $t_i$ for which the traps are kept at the distance $a_{\text{min}}$ is chosen such that alternately a $\pi$ pulse 
and a $\pi/2$ pulse are applied. The figure shows the population of the traps after $t=10,\,20$, and $25$ steps.
In Fig.~\ref{fig_1d_gs} (i) the characteristic shapes of
the quantum walk distributions are visible. 
Subsequently we will analyse how the probability distribution changes if different vibrational states are involved or
experimental imperfections are present.

\subsubsection{Excited vibrational states: the influence of temperature}

\begin{figure*}[thp]
\begin{center}
%\includegraphics[width=13cm]{fig_1d_gs_c3}\\
%\vspace{-0.9cm}
%\includegraphics[width=13cm]{fig_1d_gs_c2}\\
%\vspace{-0.9cm}
%\includegraphics[width=13cm]{fig_1d_gs_c1}
\includegraphics[width=0.8\linewidth]{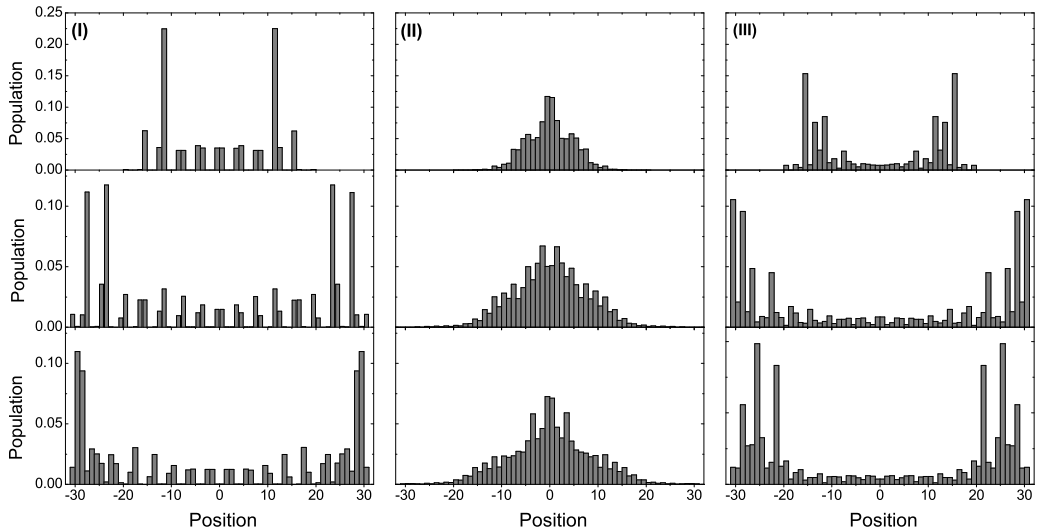}
\end{center}
\vspace{-0.9cm}
\caption[]{
The probability distribution to find the atom at a specific trap site,
with (I) the ground state and (II) the first excited state as the initial vibrational
state, for a one-dimensional quantum walk on a finite line of 62 traps; from top to bottom distributions after $t=10$, $20$, and $30$ steps are shown.
Parameters: $V_0=200\hbar\omega_x$, $\alpha a_{\text{max}}=60$, $\alpha a_{\text{min}}=28.8$, and
$\omega_x t_r=100$; $\omega_x t_{i,\pi}=20.25$ for the $\pi$ pulse and $\omega_x t_{i,\pi/2}=112$ for the $\pi/2$ pulse
(for simplicity we fixed $a_{\text{min}}$ and then searched for the smallest $t_i$ that produces the desired operation. As tunneling already happens for $a>a_{\text{min}}$,
$t_i=0$ does not give the identity operation, and for this reason $t_{i,\pi/2}>t_{i,\pi}$).
(III) like (II), but with $\omega_x t_r=200$, such that non-adiabatic excitations are suppressed.
%The distributions on the right shows the respective probabilities to detect the atom in the first excited state
%in case (ii).
}
\label{fig_1d_gs}
\end{figure*}
Tunneling as well as adiabaticity do crucially depend on the timing of the change of the trap separation. For all simulations $t_r$,
the time needed to move the traps together or apart,
and $t_i$, the time for which the trap separation is kept constant, are chosen to apply the correct operations for the vibrational ground state.
If the atom starts in an excited vibrational state, then the tunneling rate is larger and thus in general
the coin operator $C$ as well as the displacement operator $O_{\text{1D}}$ change. The former will be distinct
from the Hadamard operator $\Hop$ and in general biased,
\begin{equation}
C'\Ket{k,\pm}=\sqrt{p}\Ket{k,+}\pm\sqrt{1-p}\;e^{i\Delta_C}\Ket{k,-},\label{eq:Hmodified}
\end{equation}
(the standard unbiased Hadamard operator has $p=\frac12$ and $\Delta_C=0$), the latter will take a general form 
\begin{equation}
O'_{\text{1D}}\Ket{k,\pm}=\sqrt{c}\Ket{k\pm1,\mp}\pm\sqrt{1-c}\;e^{i\Delta_O}\Ket{k,\pm}.\label{eq:o1dmodified}
\end{equation}
($c=1$ and $\Delta_O=0$ for the standard displacement operator).
For an atom in a fixed vibrational level, $p,\,\Delta_C,\,c$ and $\Delta_O$, and thus the operators $C'$ and $O'_{1D}$
are constant, because the movement of traps is assumed to be unchanged throughout the process. In such a case 
the qualitative shape of the probability distribution is not modified significantly, it still shows the
characteristic symmetrically displaced peaks.
However, a simulation for an atom initially in the first excited vibrational state, c.f., Fig.~\ref{fig_1d_gs} (ii), shows
a distribution which essentially has a central peak and long symmetric tails. In this case the variance $\sigma^2$
grows only linearly with the number of steps, as compared to a quadratic increase of the variance 
for the ground state distribution.
The difference to the expected result can be attributed to the fact that the approaching and separating processes 
were optimized to suppress non-adiabatic excitations {\it from the ground state}. For higher vibrational states
excitations are non-negligible, causing coin as well as displacement operator to induce
transitions between different trapping states. Then effectively we have a one-dimensional walk with a higher dimensional 
coin. 
A quantum walk distribution should be reobtained when
restricting the quantum walk to some fixed vibrational state by suppressing non-adiabatic transitions.
This can be done by increasing the time $t_r$ used to approach the traps. Then, 
as can be seen in Fig.\ \ref{fig_1d_gs} (iii), again the characteristic displaced peaks of the
quantum walk probability distribution appear. As for 
the ground state, the variance increases quadratically with the number of steps.
Note however that the distribution is not the same as for the ground state due to
different coin and displacement operators.

\begin{figure}[thp]
\begin{center}
\includegraphics[width=0.8\columnwidth]{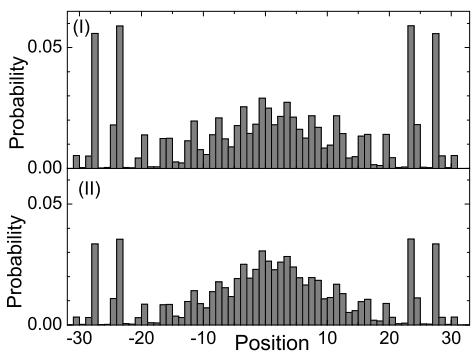}\\
\end{center}
\vspace{-0.8cm}
\caption[]{
Probability distributions after $t=20$ steps for thermal Boltzmann distributions of vibrational modes, initial ground state population
(I) $50$\% and (II) $25$\%. All other parameters as in Fig.~\ref{fig_1d_gs}.
}
\label{fig_temperature}
\end{figure}

A more realistic assumption than starting from a pure state with the atom being in a specific vibrational level is to consider a
thermal Boltzmann distribution of the vibrational modes
\begin{equation}
\rho=\frac1{z}\sum_{j=0}^{\infty}e^{-\beta E_j}\Ket{j}\Bra{j},\ \  z=\sum_j^{\infty}e^{-\beta E_j},\ \ \beta=\frac{1}{k_BT},
\end{equation}
where $E_j$ is the energy of the $j$th vibrational mode. In this case 
the experimentally accessible probability distributions are the classically averaged probability distributions, weighted with factors
$\exp(-\beta E_J)/z$. Respective probability distributions after $n=20$ steps are shown in Fig.~\ref{fig_temperature} for initial
ground state populations of $50$\% and $25$\%, corresponding to a mean number of vibrational quanta of $\ew{\nu}=1$ and
$\ew{\nu}=3$, or to a temperature of $T=1.1\;\mu$K and $T=2.7\;\mu$K (for Rb atoms and trap frequency $\omega_x=10^5$ s${}^{-1}$), respectively.
The characteristics of the quantum walk remain visible even at such $\ew{\nu}$.
In optical lattices with parameters similar to what we consider here, ground state
populations of above $98$\% have been achieved \cite{jessen}. Thus we can expect that the range of temperatures necessary
to observe the quantum distribution is well within the reach of experiments.

\subsubsection{Experimental imperfections and decoherence}

As an important experimental imperfection we will analyze 
shaking of the centers of the traps. We assume
that the movement of traps in the even or odd sets (traps $2k$ or $2k+1$, respectively) is correlated. For an implementation
with optical microtraps this is justified, because each set can be generated from a single laser beam, as described in section
\ref{expsetup}, but it will also describe the situation  for optical lattices with tunneling controlled by changing the intensity of one out of two
counterpropagating laser beams, as in \cite{vibrational,weitz1}.
We also anticipate shaking with a fixed frequency far away from the trapping frequency. This will not eliminate
transitions between vibrational states, as it alters the (optimized) path to approach the traps. 
Due to the strong sensitivity of tunneling on the distance, shaking will give rise to changes in the rate of the population oscillation
between the traps, i.e., coin and displacement operators will change. They will even be different from step to step, as shaking is
not correlated to the global motion of approaching the traps. The consequences of this random variation should be similar
to the effects observed in the presence of decoherence. Decoherence in quantum walks has been
studied by Kendon and Tregenna in a general framework \cite{QWdecoh} and by D{\"u}r {\it et al.} for the 
special case of an optical lattice implementation of a quantum walk \cite{QWOptLat}. In the presence of
decoherence, the probability distribution of the quantum walk on the (infinite) line ultimately
collapses to the gaussian distribution characteristic for the classical random walk. However, 
for the product of the number of steps $t$ and the decoherence rate $p$ being small enough, 
decoherence does not significantly degrade the quadratic spreading of the walk. As has furthermore been
found in \cite{QWdecoh}, for a certain intermediate choice of $tp$ a highly uniform distribution 
of the probabilities between positions $\pm t/\sqrt2$ can be observed if decoherence acts on the
position degree of freedom or on both, the coin and the position degree of freedom;
note that for our implementation the effect of shaking corresponds to the latter case.

\begin{figure}[thp]
\begin{center}
\includegraphics[width=\columnwidth]{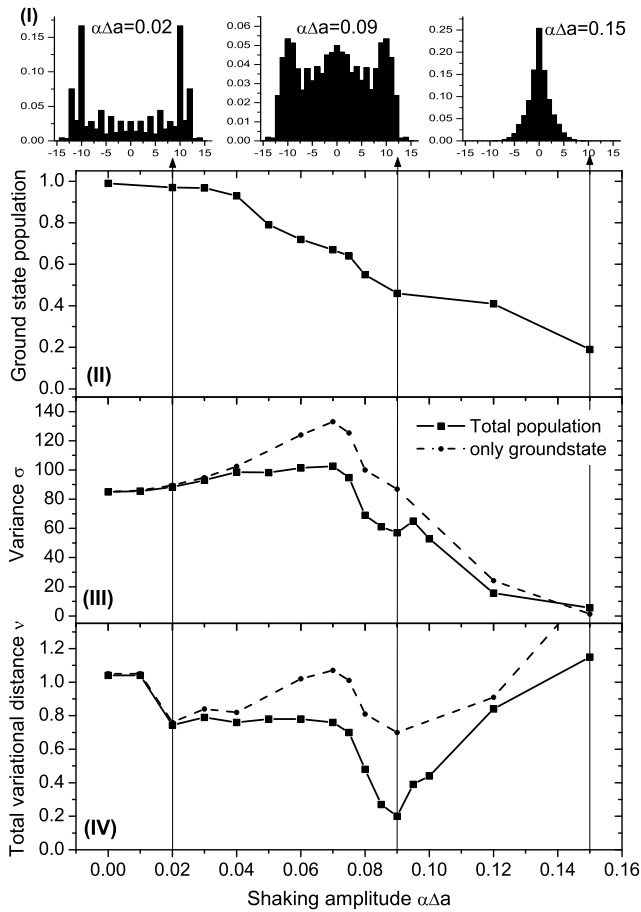}
\end{center}
\caption[]{
The effect of shaking of the trap position on the quantum walk. Shaking is modeled by a sinusoidal variation of the trap distances
around the perfect value, with frequency $\omega_{Shake}=0.01\omega_x$ and amplitude $\alpha\Delta a$
(all the other parameters are as in Fig.~\ref{fig_1d_gs}). (I) The probability
distributions for various values of $\alpha\Delta a$ (after tracing over the coin degree of freedom). (II) Ground state
population, (III) variance, and (IV) variational distance $\nu$ from the uniform distribution after $t=17$ steps.
In (II) and (III) dashed lines and squares give the respective values for the full population, dotted lines and circles for the ground state only.
}
\label{fig_shake}
\end{figure}

In Fig.\ \ref{fig_shake} the results for a sinusoidal variation of the trap distances around the perfect value with
frequency $\omega_{\text{Shake}}=0.01\;\omega_x$ and amplitude $\alpha\Delta a$ are shown.
The transition from the quantum to a classical distribution takes place for amplitudes
on the order of a percent of the minimal distance, the intermediate flat distribution is clearly visible at 
$\alpha\Delta a\approx 0.09$.
For larger amplitudes of shaking the non-adiabatic transitions are dominant (Fig.~\ref{fig_shake} (II)),
%, which produces a similar effect as for excited states (Fig.~\ref{fig_shake} (b)), i.e.~
and the variance decreases strongly with increasing shaking
(Fig.~\ref{fig_shake} (III)). For smaller amplitudes however the variance initially increases with increasing
amplitude of shaking.
%This is not remarkable on its own (e.g.~the variance will already grow if the coin operation is 
%biased to move the walker always away from the center, with ballistic motion as the extreme case)
To quantify the flatness of the distribution we also calculate the total variational distance $\nu(t)=\sum_n|P(n,t)-P_u(t)|$
to the uniform distribution $P_u(t)$ of half width $t/\sqrt{2}$ \cite{QWdecoh}. Here $P(n,t)$ is the probability to
find the particle in the traps belonging to the $n$-th qubit after $t$ steps. As Fig. \ref{fig_shake} (IV) shows, the total variational
distance decreases initially, before it increases again as the probability distribution approaches a gaussian.
%That under moderate decoherence the total variational distance can decrease, was first observed by Kendon and Tragenna
%\cite{QWdecoh}, and here we find a clear sign of the same effect, caused by an interplay of non-adiabatic transitions and
%the distribution of gate operations around a perfect operation. If only the population of the ground state is considered
%(dotted lines and circles in Figs.~\ref{fig_shake}(c) and (d)), then the the variance $\sigma$ increases even stronger,
%suggesting that the coin operation is slightly biased, and the decrease of total variational distance is weaker.

% \subsubsection{Decoherence}
Let us discuss the duration of the operations necessary for the quantum walk in order to estimate the influence of other decoherence
mechanisms in the experiment.
As the processes rely on tunneling, the duration of a single operation is on the order of the inverse trapping frequency,
which typically is about $\omega_x=10^5$ s$^{-1}$ \cite{PRLBirkl,BirklOC}. For the parameters used here a single
application of $O_{\text{1D}}\;(\eins\otimes C)$ takes around $5$ ms. The
dominant decoherence mechanism can be expected to be the scattering of photons from the trapping laser, with scattering
rates on the order of $0.1\ldots1$ s$^{-1}$ \cite{PRLBirkl,BirklOC}. Then the probability for a decoherence event to occur within a single step is
$p=0.0005\ldots0.005$ and for $t=17$ applications of $O_{\text{1D}}\;(\eins\otimes C)$ we have $tp=0.0085\ldots0.085$.
For optical lattices the same decoherence mechanism is present, but also decoherence
through fluctuations in the phase of the lasers producing the lattice, giving rise to
fluctuations in the trapping potentials should be taken into account. One the other hand operations can
be an order of magnitude faster, as the initial separation of the atoms can be made shorter, such that
decoherence rates similar to the case of optical microtraps can be expected. 
In both cases these decoherence mechanisms affect the coin as well as the position, and for
this case the crossover from the quantum to the classical distribution has been numerically
estimated in \cite{QWdecoh} to take place at $tp\approx2.6$. For this reason it should be possible
to observe the quantum walk in such systems and to analyze changes caused by temperature and shaking without 
being limited by  decoherence from photon scattering etc.
The strong dependence on temperature and on non-adiabatic transitions 
of the quantum walk with delocalized qubits might thus be interesting as a tool to analyze to which
extent the ground state population, the shaping of the trapping potentials, and tunneling processes can be controlled for a particular experimental setup.
In addition, quantum walks in this particular physical system could be used to investigate how decoherence
acts with respect to the spatial degree of freedom.

\section{A two-dimensional quantum walk}

\setcounter{subsubsection}{0}

For the two-dimensional quantum walk on a regular square lattice, i.e., if 
$$\hil_W=\Span{\Ket{(k,l)},k\,\;\text{and}\,\;l\in\{\ldots,-2,-1,0,1,2,\ldots\}},$$ 
a four dimensional coin degree of freedom to control the displacement
of the particle into the four possible directions is necessary: 
\begin{eqnarray}
\hil_C&=&\text{span}\left\{\Ket{++}\equiv\Ket{\nearrow},\Ket{-+}\equiv\Ket{\nwarrow},\right.\nonumber\\
&&\left.\quad\Ket{--}\equiv\Ket{\swarrow},\Ket{+-}\equiv\Ket{\swarrow}\right\}.\label{eqn:hilc2d}
\end{eqnarray}
Here we propose to implement such a coin by a suitable
combination of a spatially delocalized (SD) qubit and a hyperfine (HF) qubit combined with spin-dependent transport \cite{QWOptLat,spintransport},
i.e., $\hil_ {C}$ is a tensor product of the Hilbert space formed from the ground states of two adjacent traps and from two hyperfine states of the
atom: $\hil_{C}=\hil_{\text{SD}}\otimes\hil_{\text{HF}}$.

\subsubsection{Separable walk}

\begin{figure*}[t]
\begin{center}
%%%\includegraphics[width=0.4\columnwidth]{fig_2d_H_bw}
%%%\\\vspace{0.5cm}
%(IV)\\
%%%\includegraphics[width=\columnwidth]{fig_2d_O_bw}
\end{center}
\includegraphics[width=1.5\columnwidth]{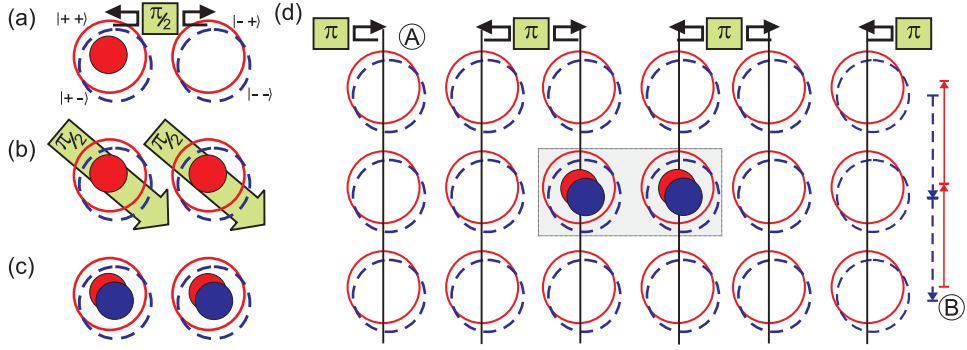}
\caption[]{(Color online)
Implementation of the coin and the coin operator $C_{\text{2D}}=\Hop_{\text{SD}} \otimes \Hop_{\text{HF}}$ for the two-dimensional walk:
(a) The four levels are formed as a tensor product of a delocalized qubit (left and right traps) and a hyperfine qubit (dark and grey filled circles symbolize the $\Ket{+}$ and
$\Ket{-}$ hyperfine states, full and dashed lines denote the respective trapping potentials).
The initial state is $\Ket{++}$. (a)$\rightarrow$(b)
For $\Hop_{\text{SD}}$ the traps are approached (for both hyperfine states). (b)$\rightarrow$(c) For $\Hop_{\text{HF}}$
a  $\pi/2$ pulse between the two hyperfine levels is applied to all traps simultaneously.
(d) The
implementation of the two-dimensional walking operator $O_{\text{2D}}$ as a combination of tunneling and spin dependent transport:
(A) traps in horizontally adjacent traps are approached to give a $\pi$-pulse as in the one-dimensional walk;
(B) the lattice is displaced in opposite vertical
directions for the two hyperfine states.
}
\label{fig_2D_H}
\label{fig_2D_O}
\end{figure*}
There is no unique extension of the Hadamard operator $\Hop$ to $\hil_C$ from Eq. (\ref{eqn:hilc2d}),
because different classes of unbiased coin operators for two-dimensional walks exist \cite{coinsandstates}. The most obvious and simple generalization 
is to take a Hadamard coin for both directions. This can be realized by first approaching the traps to perform
a $\pi/2$ pulse for the delocalized qubit as above, and then putting the atom in a superposition of the two hyperfine levels by a $\pi/2$ two-photon \cite{twophoton}
or microwave \cite{QWOptLat} pulse, which realizes $C_{\text{2D}}=\Hop_{\text{HF}}\otimes \Hop_{\text{SC}}$ 
(c.f.~Fig. \ref{fig_2D_H} (a-c) for the case of $\Ket{++}$ as initial state).
For the coin-dependent displacement assume that at each vertex of 
the two-dimensional grid each two traps forming a coin are aligned horizontally. Then in horizontal direction first the walking operator
$O_{\text{1D}}$ can be applied, i.e., within each row traps of neighboring qubits are approached as described above to give a $\pi$ pulse, followed
by translating the lattice potential in opposite vertical directions for each spin state, as proposed in \cite{QWOptLat} (see Fig. \ref{fig_2D_O} (d)).
In total, the action of the walking operator $O_{\text{2D}}$ in $\hil_W\otimes\hil_C$ is  given by
\begin{equation}
O_{\text{2D}}\Ket{(k,l),\pm\pm}=\Ket{(k\pm1,l\pm1),\mp\pm}.\label{o2d}
\end{equation}

\begin{figure}[p]
\begin{center}
%\begin{minipage}{0.3\textwidth}
\begin{flushleft}
(I)
\end{flushleft}
\includegraphics[width=0.65\columnwidth]{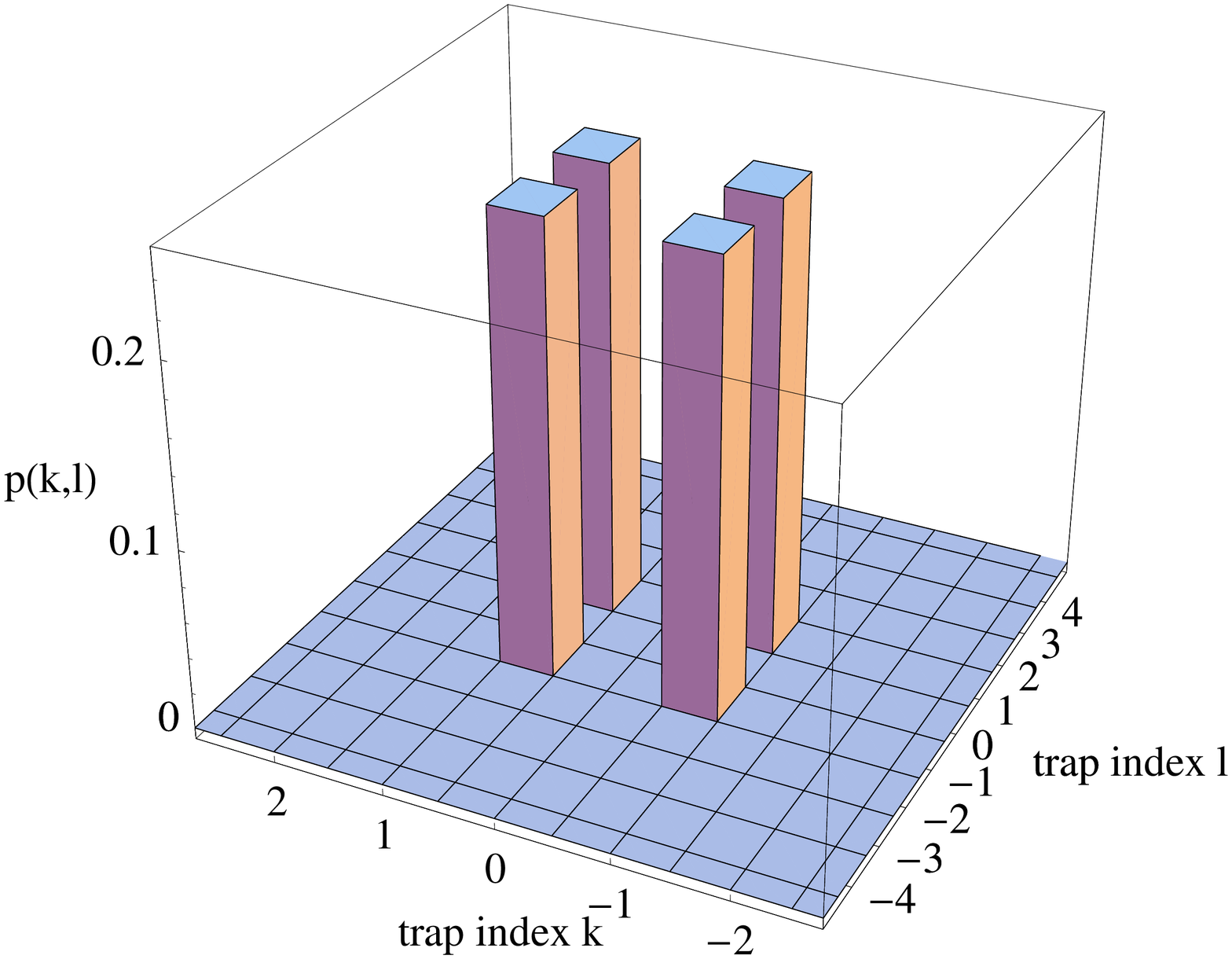}\\
\begin{flushleft}
(II)
\end{flushleft}
\includegraphics[width=0.65\columnwidth]{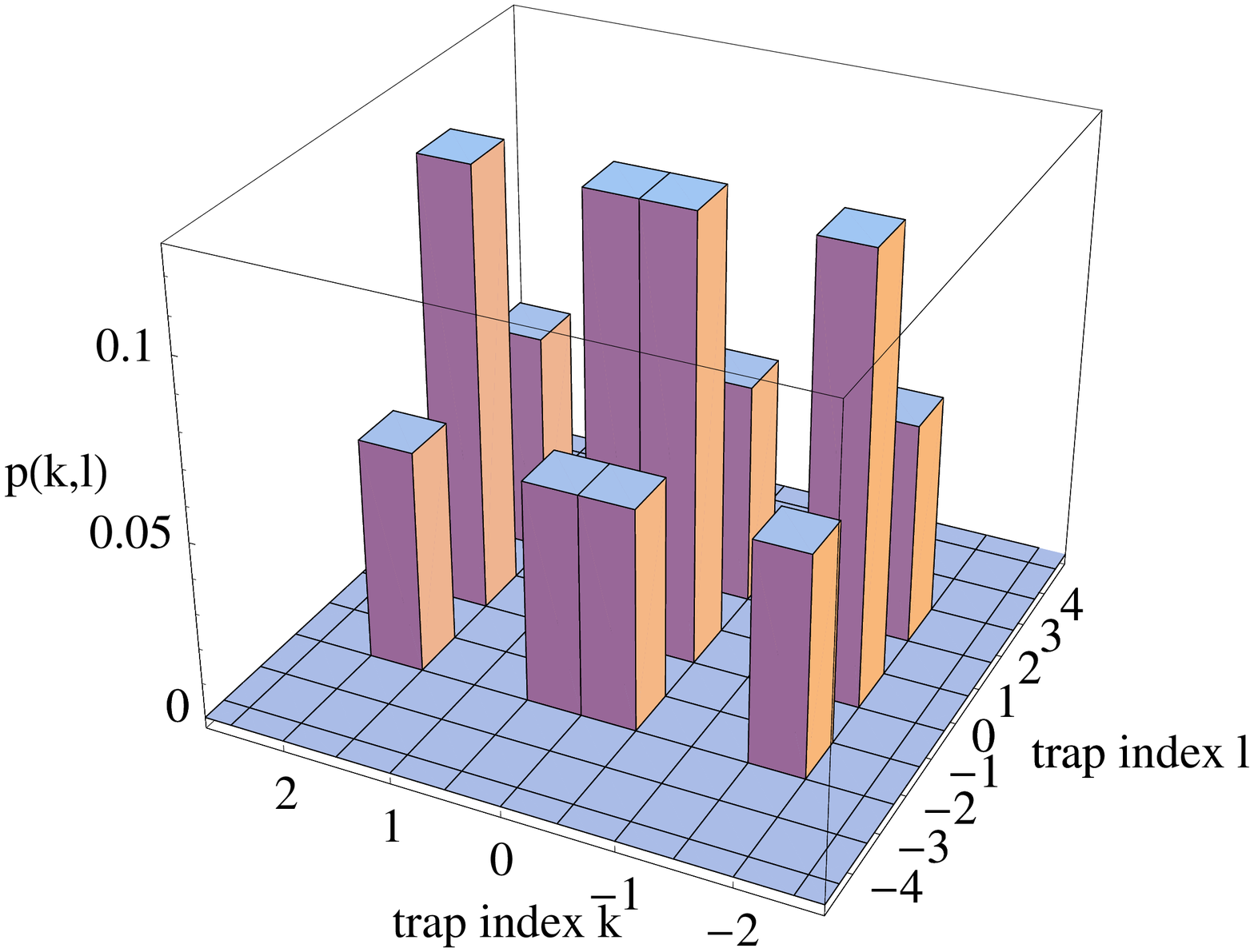}
%\end{minipage}
%\begin{minipage}{0.65\textwidth}
\begin{flushleft}
(III)
\end{flushleft}
\includegraphics[width=\columnwidth]{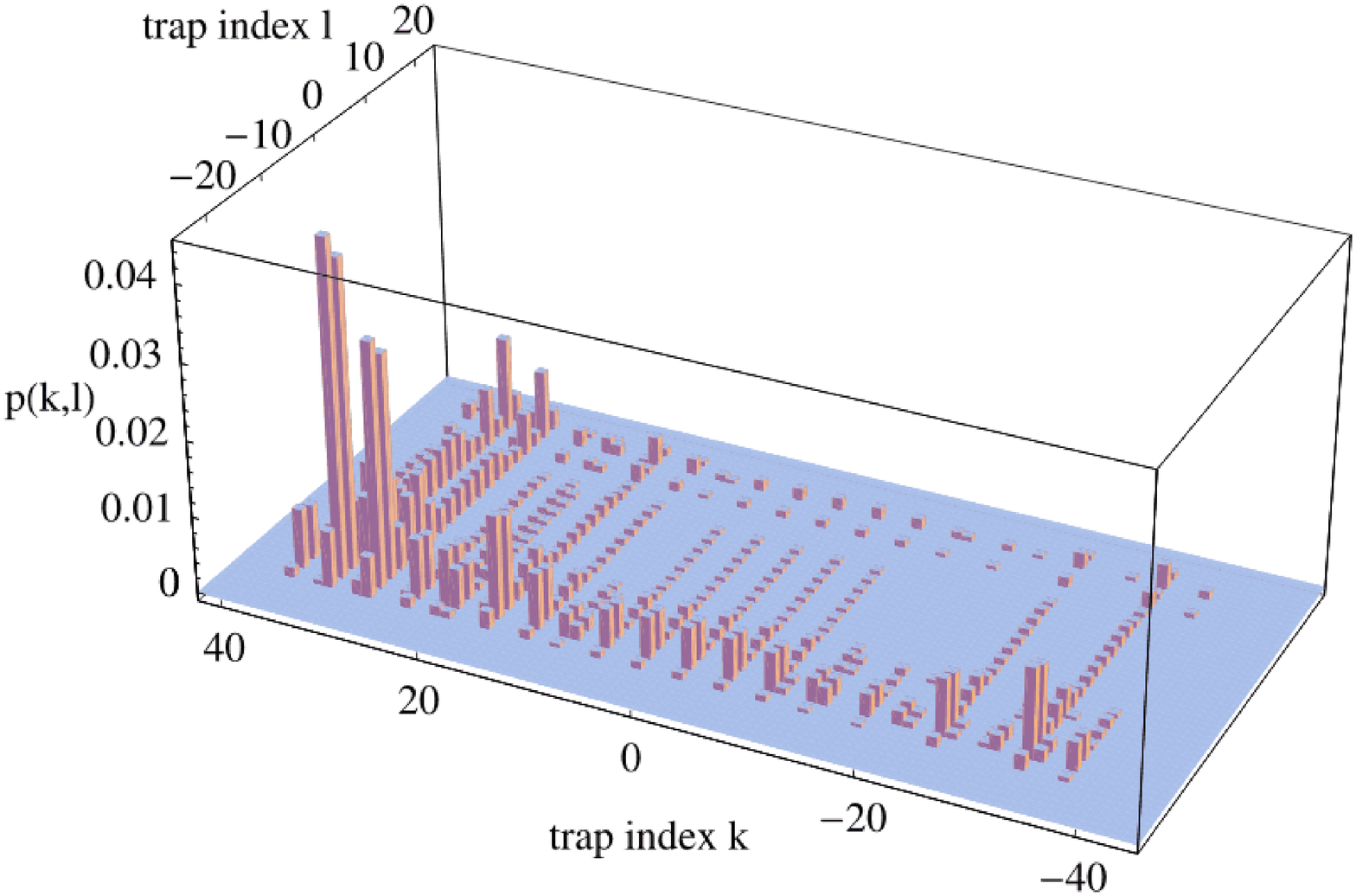}\\
%%\includegraphics[width=\columnwidth]{2d_sep_4}\\
%\end{minipage}
%\includegraphics[width=15cm]{fig_2d_prob}
\end{center}
\caption[]{
(Color online) Probability distributions for the two-dimensional quantum walk obtained from alternatingly applying 
$C_{\text{2D}}$ and $O_{\text{2D}}$ to the initial state
$\Ket{\psi_{\text{init}}}=\Ket{(0,0),++}$ 
after (I) $t=1$, (II) $t=2$, and (III) $t=25$ steps. The coordinates
on $k$ and $l$ axes label the traps, thus in k direction each two traps form one site;
note that the total population of each trap is shown (sum of probabilities for both hyperfine states). 
%The coins are oriented in the horizontal direction.
}
\label{fig_2d_prob}
\end{figure}

Fig.~\ref{fig_2d_prob} shows a probability distributions arising from alternatingly applying $\eins\otimes C_{\text{2D}}$ and $O_{\text{2D}}$ to the initial state
$\Ket{\psi_{\text{init}}}=\Ket{(0,0),++}$. From its construction it is easy to see that the coin operator $C_{\text{2D}}$
does not mix the horizontal and vertical direction. For this reason one recovers the one-dimensional quantum walk
when projecting the distributions along the $x$ or $y$ directions
(due to the choice of the initial conditions the distribution is not symmetric in this case).
$C_{\text{2D}}$ is thus a separable Hadamard walk according to the classification of \cite{QWmoreDim}.

\subsubsection{Entangled walks}

\begin{figure}[p]
\begin{center}
\includegraphics[width=\columnwidth]{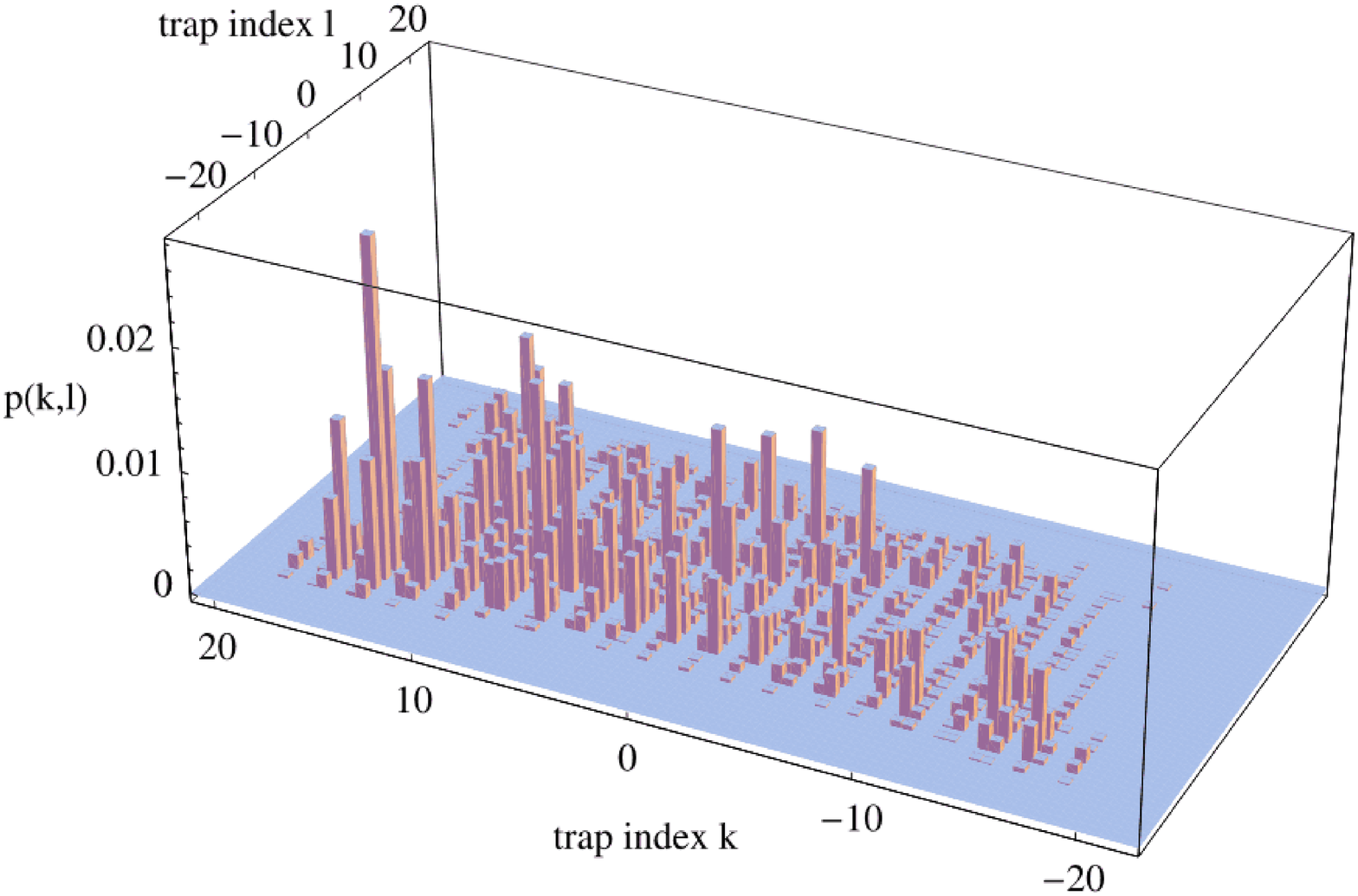}
\end{center}
\caption[]{
(Color online) Probability distribution of the two-dimensional quantum walk with the coin operator $C_{\text{2D}}^{\text{Ent}}$
after $t=25$ steps. The initial state is $\Ket{\psi_{\text{init}}}=\Ket{(0,0),++}$.
}
\label{fig_2d_ent}
\end{figure}

More sophisticated coin operators are also possible, and we will show how to implement one which entangles the two directions.
%We will show now that also more complicated coin operators are possible which entangle the two directions.
Assuming to  be able to change the trapping potentials for both hyperfine states
independently, we can apply a $\pi/2$ pulse (Hadamard operation) on delocalized qubit for the $\Ket{+}$ 
hyperfine state and a $\pi/2$ pulse followed by a $\pi$ pulse ($\NOT$ operation) on the delocalized qubit 
for the $\Ket{-}$ hyperfine state.
Subsequently a $\pi/2$ pulse is applied tothe hyperfine qubit as before. Then, defining the 
$\NOT$ operation as,
%\begin{subequations}
\begin{eqnarray}
\NOT&=&\left(
\begin{matrix}0 & 1 \\ 1 & 0\end{matrix}\right),
\end{eqnarray}
%\end{subequations}
the full coin operator reads
\begin{subequations}
\begin{eqnarray}
C_{\text{2D}}^{\text{Ent}}&=&\left(\eins\otimes \Hop^{\text{HF}}\right)
\cdot\left(
\Hop^{\text{SD}}\otimes\Ket{+}\!\Bra{+} 
+\right.\nonumber\\&&\left.+
\NOT^{\text{SD}}\Hop^{\text{SD}}\otimes\Ket{-}\!\Bra{-}
\right)\\
&=&\frac1{2}\left(
\begin{matrix}1 & 1 &1 & -1\\ 1 & -1 & 1 & 1 \\ 1 & 1 & -1 & 1 \\ 1 & -1 & -1 & -1\end{matrix}
\right).
\end{eqnarray}
\end{subequations}
(Similary, given local addressability, different laser pulses could be applied to the two traps forming the spatially delocalized
qubit).
The operator $C_{\text{2D}}^{\text{Ent}}$ is non-separable \cite{QWmoreDim,coinsandstates}.
The result
for the initial state $\Ket{\psi_{\text{init}}}=\Ket{(0,0),++}$ is shown in Fig. \ref{fig_2d_ent}.
Similar sequences of
operations can be used to generate different coin operators. It has been noted in \cite{coinsandstates} that even for a fixed
entangling coin operator very different probability distributions can be obtained through varying the initial state. At least for a
system of optical microtraps it should be possible to engineer the initial state carefully enough, because due to the large separation
of traps, single sites can easily be addressed. In this way these systems can be an interesting testbed to explore the rich structure of
probability distributions of two-dimensional quantum walks.

%\vspace{-1cm}

Recently Ambainis {\it et al.} proposed a quantum search algorithm to locate a single marked site among
$N$ locations arranged on a square $\sqrt{N}\times\sqrt{N}$
grid using a total number of $O(\sqrt{N}\log^2N)$ 
steps, thus outperforming Grovers algorithm in this case \cite{coinsfaster}, where the time
to move between locations is taken into account.
The algorithm is based on a quantum walk on a two dimensional lattice with
(i) periodic boundary conditions, (ii) a special coin as well as a special
displacement operator, and (iii) a certain initial state. We will show
how to realize the basic ingredients (ii) and (iii) 
within our proposal. The realization of periodic boundary conditions 
however is not straightforward for the setup proposed here, but might
be achievable with more advanced setups. Fixed boundaries
can be imposed for optical microtraps by only illuminating a rectangular subset of
lenses and applying an additional $\pi/2$ pulse on the hyperfine
qubits on the borders of the lattice.
Numerically we observe that the effect of a large amplitude at the marked site
is still present for such boundary conditions, though the effect
on the performance of the algorithm needs further investigation. 

The required displacement operator is the two-dimensional version of
the flip-flop walking operator which we already saw in the
one-dimensional realization. It reads
\begin{equation}
O_{\text{FF}}\Ket{(k,l),\pm\pm}=\Ket{(k\pm1,l\pm1),\mp\mp},\label{off}
\end{equation}
i.e., the walk changes direction after each step, and obviously here it can be realized by a $\pi$-pulse on the hyperfine qubit
after applying $O_{\text{2D}}$. The coin operator has to be chosen as
\begin{equation}
C_0=
\frac1{2}\left(
\begin{matrix}-1 & 1 &1 & 1\\ 1 & -1 & 1 & 1 \\ 1 & 1 & -1 & 1 \\ 1 & 1 & 1 & -1\end{matrix}
\right),
\end{equation}
except for the marked vertex, for which
\begin{equation}
C_1=
\left(
\begin{matrix}-1 & 0 & 0 & 0\\ 0 & -1 & 0 & 0 \\ 0 & 0 & -1 & 0 \\ 0 & 0 & 0 & -1\end{matrix}
\right)
\end{equation}
is applied. For an implementation of the search algorithm, these operators have to be constructed 
as a suitable combination of operations on the hyperfine and the delocalized qubit.
We will require the manipulation of the delocalized qubit to act identically on all sites, but
in order to engineer a special coin operator for the marked vertex,
for the manipulation of the hyperfine state we will assume to be able to address single
sites. % \cite{singlesite}.
We define a single qubit phase gate
%\begin{subequations}
\begin{eqnarray}
\PHASE=\left(
\begin{matrix}1 & 0 \\ 0 & -1\end{matrix}
\right),
\end{eqnarray}
%\end{subequations}
and
\begin{eqnarray}
\NOTs&=&\Ket{+}\!\Bra{+}\otimes \NOT^{\text{HF}}+\Ket{-}\!\Bra{-}\otimes\eins,
\end{eqnarray}
as an operator which produces a $\pi$ pulse only on one trap
of each coin. Then
\begin{eqnarray}
-C_0&=&\NOTs\cdot
\left(\Hop^{\text{SD}}\otimes\PHASE^{\text HF}\right)
\cdot\NOTs\cdot\nonumber\\
&&
\left(\Hop^{\text{SD}}\otimes\PHASE^{\text HF}\right)
\cdot\NOTs
\end{eqnarray}
produces, except for an overall phase, the 
correct coin operator. The operator
for the marked vertex is obtained by merely replacing
the $\NOTs$ operator by $\eins$:
\begin{eqnarray}
-C_1&=&
\left(\Hop^{\text{SD}}\otimes\PHASE^{\text HF}\right)\cdot
\left(\Hop^{\text{SD}}\otimes\PHASE^{\text HF}\right).
\end{eqnarray}
Thus, given the possibility to locally manipulate the hyperfine
qubit and except for a global phase, the quantum search coin
operators can be constructed. Finally, the initial state
has to be chosen as an eigenstate of $O_{FF}\cdot(\eins\otimes C_0)$.
For fixed boundaries, such an eigenstate is given by
$$
\Ket{\Phi_0}
=\frac1{2\sqrt{N}}\sum_{k=1}^{\sqrt{N}}\sum_{l=1}^{\sqrt{N}}\sum_{\alpha\in\{+,-\}}\sum_{\beta\in\{+,-\}}
\Ket{(k,l),\alpha\:\beta},
$$
which can be generated by a sequence of shift operations 
firstly via tunneling and secondly via displacements of the
state-dependent lattices \cite{coinsfaster}.

\section{Summary}
We have discussed the implementation of quantum walks with a neutral atom trapped in the ground state of optical
potentials by using the concept of spatially delocalized qubits, i.e., a coin defined through the presence of the atom
in one out of two trapping potentials. We have shown that in this case 
a quantum walk on a line can be performed in a simple way only through a variation of the trapping potentials,
without the need for additional lasers to address internal states of the atom. 
Our simulations were performed with realistic parameters for present optical microtrap systems, but the concept
is as well applicable to optical lattices or to magnetic microtraps.
We have studied the influences of various experimental imperfections on the probability distribution and
have found a strong change if 
%in the probability distribution if 
%in the presence of non-zero temperature 
the atom is initially not in the ground state of the trap. This change,
leading to a strong dependence of the quantum walk on temperature,
can be attributed to non-adiabatic excitations to other vibrational states during the movement of the traps.
We have also studied the influence of shaking and found a transition from quantum to classical probability
distributions, taking place for shaking amplitudes on the order of $1$\% of the tunneling distance.
As an intermediate step, this transition exhibits a very flat distribution. 
An estimate of other decoherence effects such as scattering of photons from the trapping lasers suggests 
that quantum walks should be observable in the experiment and the effects of temperature and shaking
should be accessible to experimental investigation.
In this way, implementing the quantum walk with spatially delocalized coins could give information on the
extend to which the ground state population and the movement of the traps can be controlled.

Finally, we have combined the concept of the spatially delocalized qubit with a hyperfine qubit and state dependent
potentials to obtain a scheme to implement a quantum walk on a two-dimensional regular lattice. Within this scheme,
which again is close to what is realizable with state-of-the-art technology in optical microtraps as well as in
optical lattices, different coin operators are possible, such that in this setup
the variety of different distributions in two-dimensional quantum walks can be explored.
Especially we have shown how to construct separable and entangling coin operators, as well as the operators
necessary to implement a spatial search algorithm on a two-dimensional grid.  It is worth stressing that
the scheme proposed by us can be used to construct the generalized coined quantum walk, in which the walker acquires 
at each step a phase \cite{phase}.

\section{Acknowledgments}

This work has been supported by the European Community through the HPC-Europe program, by the 
European Commission through IST projects EQUIP and ACQP,
by the DFG (Schwerpunktprogramm 'Quanteninformationsverarbeitung' and SFB 407) and
from the MCyT (Spanish Government) and the DGR (Catalan Government) under contract
BFM2002-04369-C04-02 and 2001SGR00187, respectively.
We thank D.~Bru{\ss}, R.~Corbal{\'a}n, R.~Dumke, W.~Ertmer, A. Lengwenus, T.~M\"uther, U.~Poulsen, A.~Sanpera, and M.~Volk for discussions.

\end{document}